\input{aipcheck}

\documentclass[
    ,final            
  ]
  {aipproc}

\layoutstyle{8x11single}

\begin{document}

\title{High Energy Variability Of Synchrotron-Self Compton Emitting Sources: Why One Zone Models Do Not Work And How We Can Fix It
}


\author{Philip B.  Graff}{
  address={Department of Physics, University of Maryland, Baltimore County}
}

\author{Markos Georganopoulos}{
  address={Department of Physics, University of Maryland, Baltimore County},
  altaddress={NASA, Goddard Space Flight Center}
}

\author{Eric S. Perlman}{
  address={Department of Physics and Space Sciences, Florida Institute of Technology}
  }

\author{Demosthenes Kazanas}{
address={NASA, Goddard Space Flight Center}
}

\begin{abstract} With the anticipated launch of GLAST, the existing X-ray telescopes, and the enhanced capabilities of the new generation of TeV telescopes, developing tools for modeling the variability of high energy sources such as blazars is becoming a high priority. We point out the serious, innate problems one zone synchrotron-self Compton models have in simulating high energy variability. We then present the first steps toward a multi zone model where non-local, time delayed Synchrotron-self Compton electron energy losses are taken into account. By introducing only one additional parameter, the length of the system, our code can simulate variability properly  at Compton dominated stages, a situation typical of flaring systems.
  As a first application, we were able to reproduce variability similar to that observed in the case of the puzzling `orphan' TeV flares that are not accompanied by a  corresponding X-ray flare. 

\end{abstract}

\maketitle


\section{Introduction}

In blazars, radio loud active galaxies with their relativistic jets pointing close to the line of sight,
the observed radiation is dominated by  relativistically beamed emission from the sub-pc base 
of the jet. Due to the small size of the emitting region and the large distance of such sources, currently it is not possible to spatially resolve the emitting region. Because of this, we can only obtain
information about its structure through  multiwavelength variability studies. When modeling blazar radiation, past models have mostly incorporated some form of a one zone model. Such models are limited because high-energy electrons cool faster than the source light crossing time and also because time delay effects must be considered.
When neither of these are accounted for, the model produces variability on timescales less than the light crossing time, which is incorrect. One must consider the time that it takes for light to be transmitted from different parts of the source, and as Chiaberge and Ghisellini (1999) showed, the shortest observable variability that can be trusted is the light crossing time of the zone. The basic limitation of one zone models stems from the fact that, by construction,  all the high energy variability is produced by electrons with cooling times faster than the light crossing time. One, therefore, cannot use one zone models to infer the source structure from high energy variability. 

\section{Model setup}
Kirk et al. (1998) developed an analytical model in which electrons, after being accelerated, propagate in a pipe geometry and cool through synchrotron radiation.  This results in a frequency dependent source size, in agreement with the fact that the variability timescale of synchrotron radiation increases with decreasing frequency.  Here we present a numerical code that in addition incorporates the important  inverse Compton (IC) process, and in particular the most complicated synchrotron-self Compton (SSC) losses and emissivity. We assume that a power law of relativistic electrons is injected at the base of a flow, and that the electrons flow downstream and cool radiatively. Variations in the injected plasma parameters propagate downstream and manifest themselves as frequency dependent variability.

Synchrotron and IC losses from photons external to the source are local processes, in the sense that at a given point in the flow, the energy loss rate only depends on the magnetic field and external photon field 
energy densities, and not on the conditions throughout the  source.  This is not the case with SSC losses, because synchrotron photons produced throughout the source in past times - to take into account the light travel time from one point of the source to another - contribute to the photon energy density responsible for the SSC losses and emissivity at a given point and time in the source. To incorporate this, we record  the synchrotron emissivity throughout the source as a function of time.

\section{Orphan TeV flares}

In most cases, TeV and X-ray variability are correlated (e.g. Fossati et al. 2000). A variability pattern that cannot be explained by one zone models is the so-called orphan flares, TeV flares  that are not accompanied by X-ray  flares (e.g. Krawczynski et al. 2004,  Blazejowski et. al. 2005). This cannot be simulated by one zone models because  the same electrons that produce X-rays through synchrotron raditation also IC-scatter lower energy photons to produce TeV emission. It follows then, that flares in these spectra would be linked.

We present (Fig. 1) two types of orphan flares produced by our model using a `pipe' flow geometry:
a TeV flare unaccompanied by an X-ray flare and a combined flare followed  by an ``echo'' flare in the TeV range. We produce both flares with the system being in or near a Compton-dominated state, consistent with observations. 
The first type of flare occurs when there is an additional injection of low-energy electrons; through synchrotron radiation, these provide additional low energy seed photons for IC radiation, resulting in the TeV flare. The  dip in X-ray radiation is due to the additional IC cooling of the high-energy electrons. 
The second type of flare occurs in the case of an increase in high-energy electron injection. The system responds with an initial flare in both X-ray and TeV energies. As the high-energy electrons travel down the jet and cool, they emit extra radiation in the IR. This travels back toward the base of the jet and allows for an increase in TeV emission by providing additional seed photons. 

\begin{figure}
  \includegraphics[height=.25\textheight]{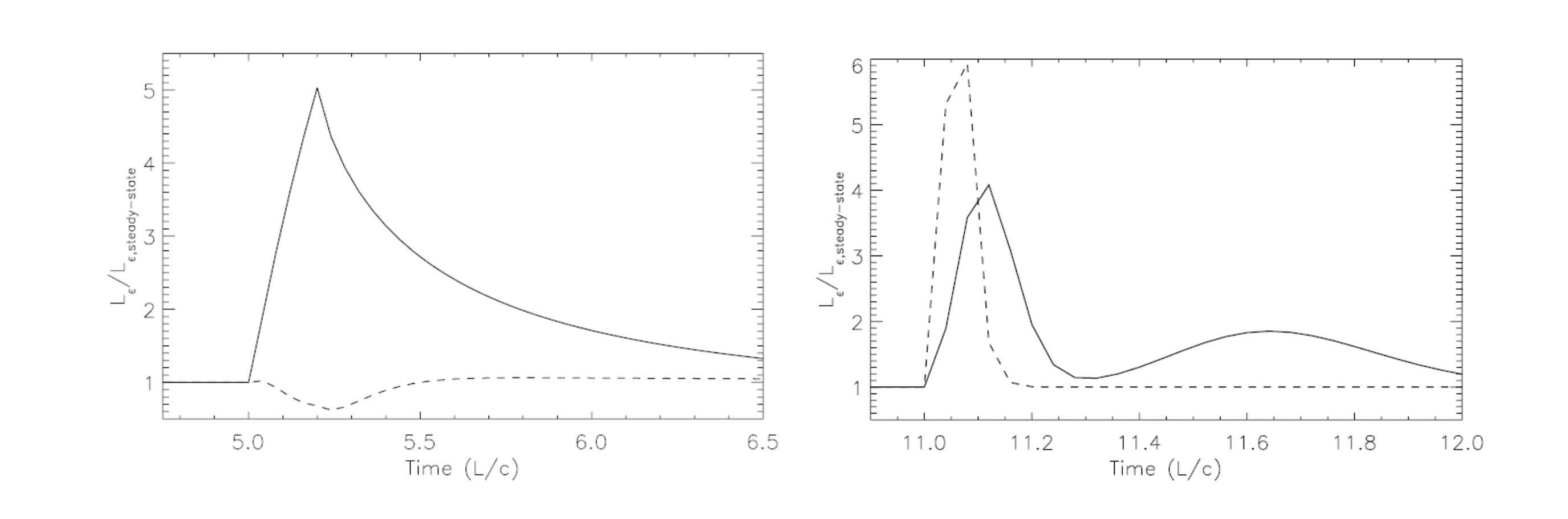}
  \caption{TeV (solid line) and X-ray (dashed line) variability from our inhomogeneous model. On the left, an event that produces a TeV flare without an accompanying X-ray flare. On the right, an event that produces an 'echo'  flare in the TeV band.}
\end{figure}

\begin{theacknowledgments}
This project is part of the senior thesis of Philip Graff,  under the supervision of Markos Georganopoulos. The authors acknowledge support from a Chandra theory grant and a NASA long-term space astrophysics grant.
\end{theacknowledgments}

\end{document}